\documentclass[12pt]{article}

\newcommand{\beq}{\begin{eqnarray}}
\newcommand{\eeq}{\end{eqnarray}}

\newcommand{\be}{\begin{equation}}
\newcommand{\ee}{\end{equation}}

\def\la{\mathrel{\mathpalette\fun <}}
\def\fun#1#2{\lower3.6pt\vbox{\baselineskip0pt\lineskip.9pt
\ialign{$\mathsurround=0pt#1\hfil ##\hfil$\crcr#2\crcr\sim\crcr}}}
\newcommand{\vex}{\mbox{\boldmath${\rm x}$}}
\newcommand{\vep}{\mbox{\boldmath${\rm p}$}}
\newcommand{\vek}{\mbox{\boldmath${\rm k}$}}
\newcommand{\lan}{\langle}
\newcommand{\ran}{\rangle}

\title{Masses and decay constants of $B_q$ mesons in the QCD string approach}

\author{A.M. Badalian\footnote{badalian@itep.ru} 
and Yu.A. Simonov\footnote{simonov@itep.ru}\\
Institute of Theoretical
and Experimental Physics, Moscow, Russia\\
B.L.G. Bakker\footnote{blg.bakker@few.vu.nl}\\
Vrije Universiteit, Amsterdam, The Netherlands}

\begin{document}

\maketitle

\begin{abstract}
The relativistic string Hamiltonian is used to calculate the masses and
decay constants of $B_q$ mesons: they appear to be expressed through
onlythree fundamental values: the string tension $\sigma $,
$\alpha_s$,  and  the quark pole masses. The values $f_B =186 $ MeV,
$f_{B_s}= 222$ MeV are calculated while $f_{B_c}$ depends on the
$c$-quark pole mass used, namely $f_{B_c}=440~(424)$ MeV for $m_c
=1.40~(1.35)$ GeV. For the $1P$ states we predict the spin-averaged
masses:  $\bar M(B_J)=5730$ MeV and $\bar M(B_{sJ})=5830$ MeV which are
in good agreement with the recent data of the D0 and CDF
Collaborations,  at the same time owning to the string correction being
by $\sim 50$ MeV smaller than in other calculations.

\end{abstract}

\section{Introduction}

The decay constants of pseudoscalar  (P) mesons $f_P$ can be directly
measured in $P\to \mu \nu$ decays \cite{1} and therefore they can be
used as an important criterium to compare different theoretical
approaches and estimate their accuracy. Although during the last decade
$f_P$  were  calculated many times: in potential models \cite{2,3,4},
the QCD sum rule method \cite{5}, and in lattice QCD \cite{6,7}, here
we again address the properties of the $B, B_s, B_c$ mesons for several
reasons.

First, we use here the relativistic string Hamiltonian (RSH)
\cite{8}, which is derived from the QCD Lagrangian with the use of the
field correlator method (FCM) \cite{9} and successfully applied to
light mesons and heavy quarkonia \cite{10,11}. Here we show that
the meson Green's function and decay constants can also be derived
with the use of FCM.

Second, the remarkable feature of the RSH $H_R$  and also the
correlator of the currents $G(x)$ is that they are fully determined by
a minimal number of fundamental  parameters:  the string tension
$\sigma$, $\Lambda_{\overline{MS}}(n_f)$, and the pole (current) quark
masses $m_q(\bar m_q)$. All these parameters are taken to be fixed from
our analysis of heavy quarkonium spectra \cite{10} and  light meson
Regge trajectories \cite{11}:
\be
 \sigma=0.18 {\rm ~GeV}^2; ~~\Lambda_{\overline{MS}}^{(4)}  =250(5) {\rm MeV};
\label{1}
\ee
and the pole masses taken are
\beq m_{u(d)} &=&0;~~  m_s =170 (10) ~{\rm MeV};
\nonumber \\
 m_c &=& 1.40 ~{\rm GeV};~~ m_b =4.84  ~{\rm GeV}.
\label{2}
\eeq

Third, recently new data on the masses of $B_c$ and the $P$-wave
mesons: $B_1$, $B_2$, and $B_{s2}$ have been reported by the D0
and CDF Collaborations \cite{12,13}, which give  additional information
on the $B_q$ -meson spectra. Here we calculate the spin-averaged masses
of the $P$-wave states $B$ and $B_s$.

We would like to emphasize here that in our relativistic calculations
no constituent masses are used. In the meson mass formula an overall
(fitting) constant, characteristic for potential models, is absent and
the whole scheme appears to be rigid.

Nevertheless, we take into account an important nonperturbative (NP)
self-energy contribution to the quark mass, $\Delta_{SE} (q)$ (see
below eq. (\ref{15})). For the heavy $b$ quark $\Delta_{SE} (b)=0$  and
for the $c$ quark $\Delta_{SE} (c)\simeq -20$ MeV \cite{10}, which is
also small.

For any kind of mesons we use a universal static potential with pure
scalar confining term,
\be
 V_0 (r) =\sigma r -\frac{4}{3} \frac{\alpha_B(r)}{r},
\label{3}
\ee
where  the coupling $\alpha_B(r)$ possesses  the asymptotic freedom
property and saturates at large distances with $ \alpha_{crit} (n_f=4)
=0.52$ \cite{14}. The coupling can be expressed through $\alpha_B(q)$
in momentum space, 
\be 
\alpha _B(r) =\frac{2}{\pi} \int^\infty_0
dq \frac{\sin qr}{q} \alpha_B (q),
\label{4}
\ee 
where 
\begin{equation}
 \alpha_B(q) =\frac{4\pi}{\beta_0 t_B} \left( 1-
\frac{\beta_1}{\beta_0^2} \frac{\ln t_B}{t_B}\right)
\label{eq.4a}
\end{equation}
 with $t_B=\ln
\frac{q^2+M^2_B}{\Lambda^2_B}.$ Here the QCD constant $\Lambda_B$,  is
expressed as  \cite{15} 
\be  
 \Lambda_B(n_f)= \Lambda_{ \overline{MS}}
 \exp \left\{\frac{1}{2\beta_0} \cdot \left( \frac{31}{3} -\frac{10}{9}
 n_f\right)\right\}
\label{5}
\ee 
and $M_B(\sigma, \Lambda_B)=(1.00\pm 0.05) {\rm~ GeV}$ is  the so
called background mass \cite{14}. For heavy-light mesons  with  $
\Lambda_{ \overline{MS}} (n_f=4) =250 (5) ~{\rm MeV} $  one obtains $
\Lambda_B (n_f=4) =355 (7) ~{\rm MeV}. $

\section{String Hamiltonian }

At the first stage (1993-2005) the RSH was derived and applied to
mesons, glueballs, hybrids, and  baryons\cite{9,10}. In
all cases a good agreement with experiment and lattice results
have been obtained for {\em the same minimal set of parameters}.

In this work the same method is applied to the correlator of
the currents which defines the decay constant $f_\Gamma$ in any
channel $\Gamma$. For a meson the RSH can be presented as \cite{8,9}
\be
 H=H_0+\Delta H,
\label{6}
\ee 
where  $\Delta H=V_{LS} +V_{SS}
+V_T+V_{SE}$ is treated as a perturbation and  every term  can be
derived  within  the same method. The unperturbed RSH  was
deduced in \cite{8}: 
\be 
H_0 =\sum_{i=1,2} \left(
\frac{\omega_i}{2} + \frac{m_i^2+\vep^2}{2\omega_i}\right) + V_0
(r);
\label{7}
\ee 
\be 
H_0\varphi_n =M_n\varphi_n.
\label{8}
\ee 
In
(\ref{7})  $m_1(m_2)$ is the pole (current) mass of a quark
(antiquark). The variable $\omega_i$ is defined from extremum
condition, which is  taken  either from\\
(1) The exact condition: $\frac{\partial H_0}{\partial\omega_i} =0$,
 which gives
\be
 \omega_i =\sqrt{\vep^2+m^2_i}.
\label{9}
\ee 
Then 
\be 
H_0\varphi_n =\left\{\sqrt{\vep^2+m^2_1}+\sqrt{\vep^2+m^2_2}+
    V_0(r)\right\} \varphi_n =  M_n\varphi_n
\label{10}
\ee
reduces to the Salpeter equation, which just  defines $\omega_i{(n)}
=\lan \sqrt{\vep^2+m^2_i}\ran_n$ as a constituent mass.\\
(2)  The approximate condition: $\frac{\partial  M_n}{\partial\tilde
\omega_i} =0$ (the so-called einbein approximation). As shown  in
\cite{9} the difference between $\omega_i$ and $\tilde \omega_i$ is
$\la$~5\%.

For the RSH (\ref{6}) the spin-averaged mass $ \overline{M}(n,L)$  is
given by a simple expression: 
\beq 
\overline{M} (nL)
&=&\frac{\omega_1}{2}+\frac{\omega_b}{2} + \frac{m^2_1}{2\omega_1}
+\frac{m^2_b}{2\omega_b} +E_n \left( \mu \right)
\nonumber\\
&-&\frac{2\sigma\eta_f}{\pi\omega_1} -\Delta_{str} (L\neq 0),
\label{11}\eeq 
with  
\be
\omega_i(nL) =\lan
\sqrt{\vep^2+m_i^2}\ran_{nL}; ~~\mu
=\frac{\omega_1\omega_b}{\omega_1+\omega_b}.
\label{12}
\ee 
In the
general case, the self-energy term $\Delta_{SE}$ is shown to be
defined by the analytic formula \cite{16} 
\be
\Delta_{SE} (q_f)
=-\frac{2\sigma \eta_f}{\pi\omega_f};
\label{13}
\ee 
with $\eta_f=0.9$ for a $u(d)$ quark, $\eta_f\cong0.7$ for an $s$
quark, $\eta_f= 0.4$ for a $c$ quark, and $\eta_b=0$. Therefore, for a
$b$ quark $\Delta_{SE} (b) =0$. The mass formula (\ref{11}) does not
contain any overall constant ${C}$. Note that the  presence of $C$
violates linear  behavior  of Regge trajectories.

The calculated masses of the low-lying states of $B$, $B_s$, and $B_c$
mesons are given in Table~\ref{tab1}, as well as their  values  taken from
\cite{2,3,6,7}.

\begin{table}
\begin{center}
\caption{Masses of the low-lying $B_q$ mesons in the QCD String Approach}

\begin{tabular}{lll}\hline
 Meson  & $M(nL)$& Exp. \\\hline
  $B$ & 5280(5)$^a$ & 5279.0(5) \\
  &5310$^2$&\\
  &5275$^3$&\\
\hline
  $B^*$ & 5325$^a$& 5325.0(6) \\
  &5370$^2$&\\
  &5326$^3$&\\\hline
  $B_1(1P)$ &$\bar M= 5730^a$ &  5721(8) D0 \\
  &&5734(5) CDF\\
  \hline
    $B_2(1P)$  &$\bar M=5730^a$  &  5746(10) D0\\
 &5800$^2$&5738(6) CDF\\
  \hline
  $B_s$ & 5369$^a$ &5369.6(24) \\ &5390$^2$&\\
  &5362$^3$&\\
  \hline
  $B_s^*$ & 5416$^a$ &  \\
  &5450$^2$&5411.7(32)\\
  &5414$^3$&\\
  \hline
$  B_{s2}$ & $\bar M=5830$&  5839(3) D0 \\ &5880$^2$&\\ \hline
$ B_c$ & 6280(5)$^a$ &\\
 &6271$^2$ &6275(7) CDF \\
 &6304(12)$^6$&\\
 \hline
  $B_c^*$ &  6330(5)$^a$ &   \\
  &6338$^2$&\\
  &6321(20)$^6$&\\\hline
\multicolumn{3}{c}{$^a$ Masses are calculated in this paper.}
\\
\end{tabular}
\end{center}

\label{tab1}
\end{table}

It is of interest to notice that in our calculations the masses
of the $P$-wave states appear to be by 30-70 MeV lower than in
\cite{2} due to taking into  account a string correction
\cite{11}.

\section{Current Correlator}

The FCM can be also used to define the correlator  $G_{\Gamma} (x)$ of
the currents $j_\Gamma(x)$,
\begin{equation}
 j_\Gamma(x) =\bar \psi_1(x) \Gamma \psi_2(x),
\label{eq.13a}
\end{equation}
for $S$,$P$,$V$, and $A$ channels (here the operator $\Gamma=t^a\otimes
(1,\gamma_5, \gamma_\mu, i\gamma_\mu \gamma_5)$). The correlator,
\be  
G_\Gamma(x) \equiv \lan j_\Gamma (x) j_\Gamma (0)\ran_{\rm vac},
\label{14}
\ee 
with the use of spectral decomposition of
the currents $j_\Gamma$ and the definition,
\begin{eqnarray}
 \lan {\rm vac} |\bar \psi_1 \gamma^0 \gamma_5 \psi_2 | P_n (\vek =0)
\ran & = & f^P_n M_n, \quad (A ,~P)
\nonumber \\
 \lan {\rm vac} |\bar \psi_1 \gamma^\mu  \psi_2 | V_n (\vek,\varepsilon)
\ran & = & f^V_n M_n \varepsilon^\mu, \quad (V)
\label{eq.14a}
\end{eqnarray}
can be presented as \cite{3} 
\be 
 \int G_\Gamma(x) d\vex =\sum_n \frac{M_n}{2}[f^{P}_n]^2 e^{-M_nT}.
\label{15} 
\ee 
On the other
hand, applying the FCM and RSH, a very useful relation can  be derived
\cite{18}: 
\beq
 \int G_\Gamma(x) d\vex & = &
 \frac{N_c}{\omega_1\omega_2} \lan 0|Y_\Gamma e^{-H_0T}| 0\ran
\nonumber \\ 
 & = & \frac{N_c  Y_\Gamma}{\omega_1\omega_2}
 \sum_n |\varphi_n (0)|^2 e^{-M_nT}.
\label{16}
\eeq 
Here  $Y_P(Y_V)$ for the $P(V) $ channel is  given by a simple expression: 
\beq
 Y_P & = & m_1m_2 +\omega_1\omega_2 -\lan \vep^2\ran,
\nonumber \\
 Y_V & = & m_1m_2 +\omega_1\omega_2 + \frac13\lan \vep^2\ran.
\label{18}
\eeq
Then from  Eqs.~(\ref{15}) and (\ref{16}) one obtains the following analytical
expression for the decay constants (for a given state labelled $n$): 
\be
 \left[ f_n^{P(V)}\right]^2 =\frac{2N_c}{\omega_1\omega_b}
 \frac{ Y_\Gamma}{M_n} |\varphi_n(0)|^2.
\label{19}
\ee 
This very transparent formula contains only  well defined factors:
$\omega_1$ and $\omega_b$, the meson mass $M_n$, and $\varphi_n$ the
eigenvector of $\hat H_0$. Then in the $P$ channel
\be
 \left( f^P_n\right)^2 =\frac{6(m_1m_2+\omega_1\omega_2-\lan
 \vep^2\ran)}{ \omega_1\omega_2 M_n} |\varphi_n(0)|^2,
\label{20}
\ee
where the w.f. at the origin, $\varphi_n(0)$, is  a relativistic
one. In the nonrelativistic limit $\omega_i\to m_i,$
$\varphi_n(0)\to \varphi_n^{\rm NR}(0)$ and one comes to the standard
expression:
\begin{equation}
 \left( f^P_n (0)\right)^2 \to \frac{12}{M_n} |\varphi^{NR}_{n} (0)|^2.
\label{eq.20a}
\end{equation}

The calculated decay constants are given in Table~\ref{tab1} and their values
turn out to be in good agreement with lattice data \cite{6,7} and the
predictions of Ref.~\cite{3}. Our analysis  also shows  that $f_{B_c}$
is sensitive to the value taken for $m_c$ (pole).

\begin{table}
\begin{center}
\caption{Pseudoscalar constants of $B_q$ mesons (in MeV)}

\begin{tabular}{llll}\hline
   & EFG$^3$ & Lattice$^{6,7}$ & This paper \\ \hline
  $f_B$ & 189 & 216(34) & 186(5) \\
  $f_{B_s}$  & 218 & 249(42) & 222(2) \\
   $\frac{f_{B_s}}{f_B}$  & 1.15 & 1.20(4) & 1.19(2) \\
  $f_{B_c}$   & 433 & 420(20) & 438(8) \\
   &  & 440(20) & \\\hline
\label{tab2}
\end{tabular}
\end{center}
\end{table}

\section{ Conclusions}

From our analysis it follows that
\begin{itemize}
\item  The dynamics of heavy-light mesons is  sensitive to the number of
flavors $n_f$ and to the value of the quark pole mass used  (more
sensitive than in the case of $c\bar c$ and $b\bar b$ spectra). For $B_q$ mesons
$n_f=4$ is used.  

\item Solutions of the Salpeter equation using the RSH give the masses
of $B,~B_1$, and $B_2$ and also $B_s$ and $B_c$ in agreement with
experiment within $\pm 10$ MeV accuracy.

\item For $B^*$ and $B^*_s$ agreement with experimental values is
reached if $\alpha_{\rm HF} =0.32(1)$ s used in the hyperfine
interaction.

\item For  the same $\alpha_{HF} =0.32$ we predict $\Delta_{HF}
(B_c)=50$ MeV or $M(B^*_c)=6325$ MeV.

\item  In  our analytic approach with minimal input of fundamental
parameters ($\sigma, \alpha_{s}, m_i)$ the calculated decay constants
are $f_B=186~{\rm MeV},~f_{B_s}=222~{\rm MeV},~f_{B_s}/f_B =1.19$.

\item For $B_c$ the decay constant is very sensitive to $m_c$ (pole): 
$f_{B_c} = 440~{\rm MeV}~ (m_c=1.40~{\rm GeV})$ and $ f_{B_c} =425~{\rm
MeV}~ (m_c=1.35~{\rm GeV})$

\end{itemize}

\end{document}